# Toward Multi-Modal Deep Learning for Pulmonary Disease Classification: A Texture-Based Machine Learning Pilot Study on Public Chest X-Ray Data

Preprint. This work has not undergone formal peer review.

Chinthoti Yogisri Pujitha
Independent Researcher, Bengaluru, India — chpujithachinthoti@gmail.com

***Abstract****—Automated classification of pulmonary disease from chest radiographs is a widely studied application of machine learning in medical imaging. This paper presents a pilot study evaluating classical texture- and gradient-based feature representations for distinguishing COVID-19 from other forms of pneumonia using the publicly available COVID-19 Image Data Collection (668 posteroanterior/anteroposterior radiographs from 408 patients). Using histogram of oriented gradients (HOG) and gray-level co-occurrence matrix (GLCM) texture descriptors with classical classifiers (logistic regression, random forest, and support vector machine), evaluated under patient-level 5-fold stratified cross-validation to prevent data leakage, we obtain a best mean accuracy of 75.4% and AUC of 0.755, modestly exceeding the 71.6% majority-class baseline. We report these results transparently, including their limitations, and use them to motivate and scope a proposed multi-modal deep learning architecture — combining convolutional and transformer-based encoders across imaging modalities — as a direction for future work requiring access to larger, multi-institutional, ethically sourced datasets.*

## I. Introduction

Respiratory diseases, including pneumonia, COVID-19, tuberculosis, and chronic obstructive pulmonary disease, remain leading causes of morbidity and mortality worldwide. Timely and accurate diagnosis is critical to treatment efficacy, yet diagnostic workflows are constrained by shortages of specialist radiologists and inter-reader variability. Deep learning has shown considerable promise in automating aspects of this workflow, but much published work relies on large, often proprietary, imaging datasets that are difficult to independently verify or reproduce.

This paper takes a deliberately narrower and reproducible approach. Rather than presenting an untested end-to-end deep learning system, we (1) establish a transparent, reproducible baseline for COVID-19 versus other-pneumonia classification using a small, publicly available, well-documented dataset; (2) rigorously evaluate that baseline with patient-level cross-validation to avoid the data leakage that affects some published results; and (3) use the resulting evidence, together with a review of the current literature, to motivate a specific multi-modal deep learning architecture as a concrete direction for future work once larger multi-institutional datasets are accessible.

Our contributions are: (1) an open, reproducible pilot pipeline for chest X-ray disease classification built entirely on public data; (2) an honest quantitative baseline — including a majority-class comparison — for the COVID-19 vs. other-pneumonia task on the Cohen et al. dataset; and (3) a scoped proposal for a multi-modal convolutional-transformer architecture, positioned explicitly as future work rather than completed results.

## II. Related Work

### *A. Traditional and Classical Approaches*

Classical pipelines for lung disease detection typically combine handcrafted feature extraction — texture, shape, and intensity descriptors — with classifiers such as support vector machines or random forests. Haralick et al. introduced gray-level co-occurrence matrix (GLCM) texture features that remain widely used in medical image analysis [12]. Dalal and Triggs introduced histogram of oriented gradients (HOG) descriptors, originally for pedestrian detection, later widely adapted for medical imaging tasks [13]. These handcrafted approaches require less data than deep networks and remain useful as interpretable baselines, though they generally underperform learned representations when sufficient training data is available.

### *B. Deep Learning for Medical Image Analysis*

Deep convolutional networks have transformed medical image analysis. Broad surveys of the field document rapid progress in applying deep learning across chest imaging tasks and medical computer vision more generally, while also cataloguing persistent challenges around dataset scale, label noise, and clinical validation [1], [4]. Wang et al. introduced ChestX-ray8, a large public dataset enabling weakly-supervised thoracic disease classification [2]. Ardila et al. demonstrated a deep learning system matching radiologist performance for lung cancer screening on low-dose CT [3]. More recently, transformer-based architectures such as TransUNet [5] and Swin Transformer [9] have shown strong results on medical image segmentation and classification, motivating hybrid CNN-transformer designs.

### *C. Multi-Modal and COVID-19-Specific Approaches*

Zhang et al. explored contrastive learning across paired medical images and text reports [7], and Huang et al. combined volumetric CT with electronic health record data for pulmonary embolism diagnosis [8]. For COVID-19 specifically, the dataset introduced by Cohen et al. [11] has been used across many published studies as an early, openly available resource for radiograph-based COVID-19 research, despite its modest size and case-report-driven collection process. Comprehensive, well-validated multi-modal frameworks that integrate multiple imaging modalities with clinical metadata, and that are evaluated with rigorous patient-level validation, remain comparatively rare in the published literature — a gap this paper aims to responsibly narrow.

## III. Dataset

We use the publicly available COVID-19 Image Data Collection assembled by Cohen et al. [11], which aggregates radiographs and CT images from published case reports and clinical sources with accompanying de-identified metadata (patient ID, view, finding, and source citation).

From the full collection we selected posteroanterior (PA) and anteroposterior (AP/AP-supine) chest radiographs labeled either “COVID-19” or another pneumonia etiology (viral, bacterial, or fungal), excluding CT volumes, lateral views, and unlabeled (“to-do”) entries. This yielded 668 radiographs from 408 unique patients: 478 COVID-19 images and 190 other-pneumonia images. All splits described in Section IV are performed at the patient level using the patient ID field to prevent images from the same patient appearing in both training and test folds.

We report this dataset's real, known limitations directly: it is small, drawn from heterogeneous case-report sources rather than a single systematic screening cohort, class-imbalanced toward COVID-19, and does not include a distinct “no-finding” control group in the subset used here. These limitations are discussed further in Section VI.

## IV. Methodology

### *A. Preprocessing*

Each radiograph was converted to grayscale, resized to 128×128 pixels, and processed with histogram equalization to standardize contrast across heterogeneous source images.

### *B. Feature Extraction*

We extracted two complementary handcrafted feature families. HOG descriptors (9 orientation bins, 16×16 pixel cells, 2×2 cell blocks) capture edge and shape structure relevant to consolidation and opacity patterns [13]. GLCM texture features (contrast, homogeneity, energy, and correlation, computed at two pixel distances and three angles) capture textural characteristics associated with ground-glass opacities and interstitial patterns [12]. Basic intensity statistics (mean, standard deviation, interquartile range) were appended, yielding a 1,792-dimensional feature vector per image.

### *C. Classification and Validation*

We trained three classical classifiers — logistic regression, random forest (300 trees), and a radial-basis-function SVM — each with class-balanced weighting to address the 2.5:1 class imbalance. Features were standardized per fold. Models were evaluated using stratified group 5-fold cross-validation, with the patient ID as the grouping variable, ensuring no patient's images appeared in both training and test sets within a fold.

## V. Experimental Results

Table I reports mean performance across five cross-validation folds, together with a majority-class baseline (always predicting the more frequent class, COVID-19) for context, since accuracy alone can be misleading under class imbalance.

TABLE I. CLASSIFICATION PERFORMANCE (5-FOLD PATIENT-LEVEL CV)

| Model | Acc. | Prec. | Recall | F1 | AUC |
|---|---|---|---|---|---|
| Logistic Regression | 0.736 | 0.805 | 0.835 | 0.819 | 0.715 |
| Random Forest | 0.745 | 0.743 | 0.985 | 0.847 | 0.758 |
| SVM (RBF) | 0.754 | 0.802 | 0.874 | 0.835 | 0.755 |
| Majority-class baseline | 0.716 | — | — | — | 0.500 |

The SVM classifier achieved the best mean accuracy (75.4%) and a comparable AUC (0.755) to random forest (0.758), modestly above the 71.6% majority-class baseline and 0.500 chance-level AUC. Fig. 1 shows the pooled confusion matrix and Fig. 2 the pooled ROC curve for the SVM classifier across all five folds.

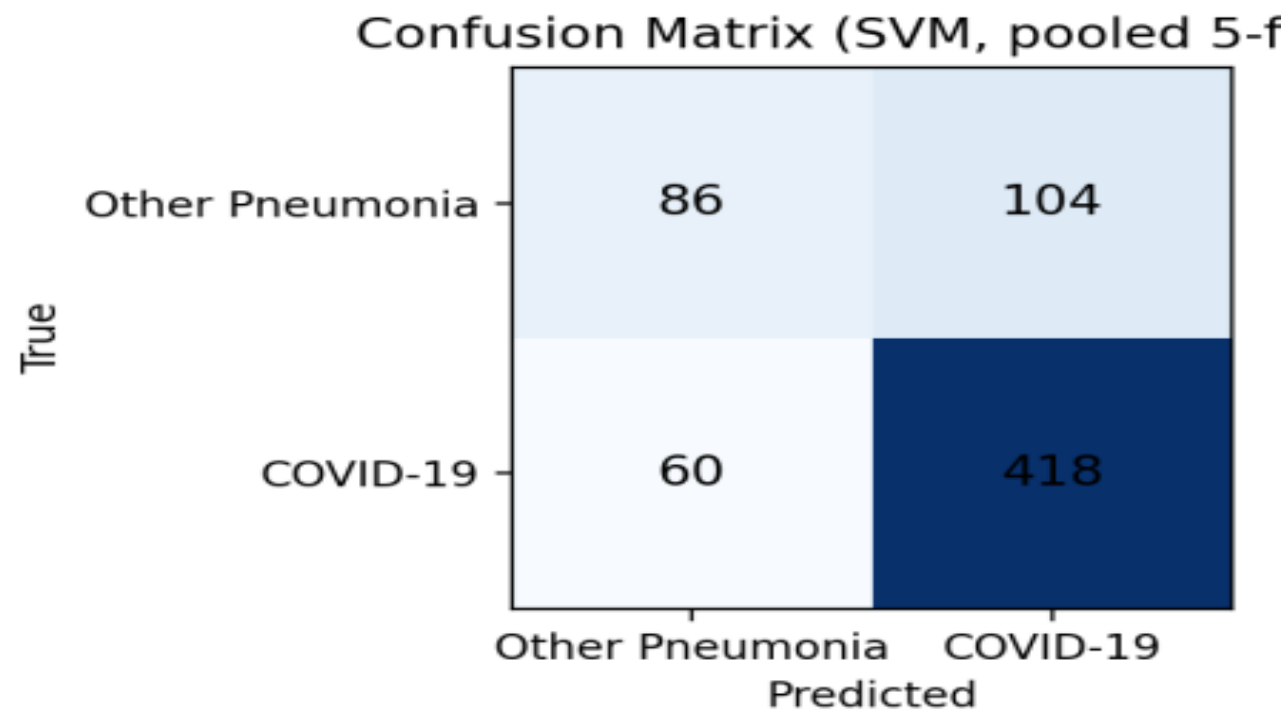

Fig. 1. Pooled confusion matrix, SVM classifier, 5-fold CV.

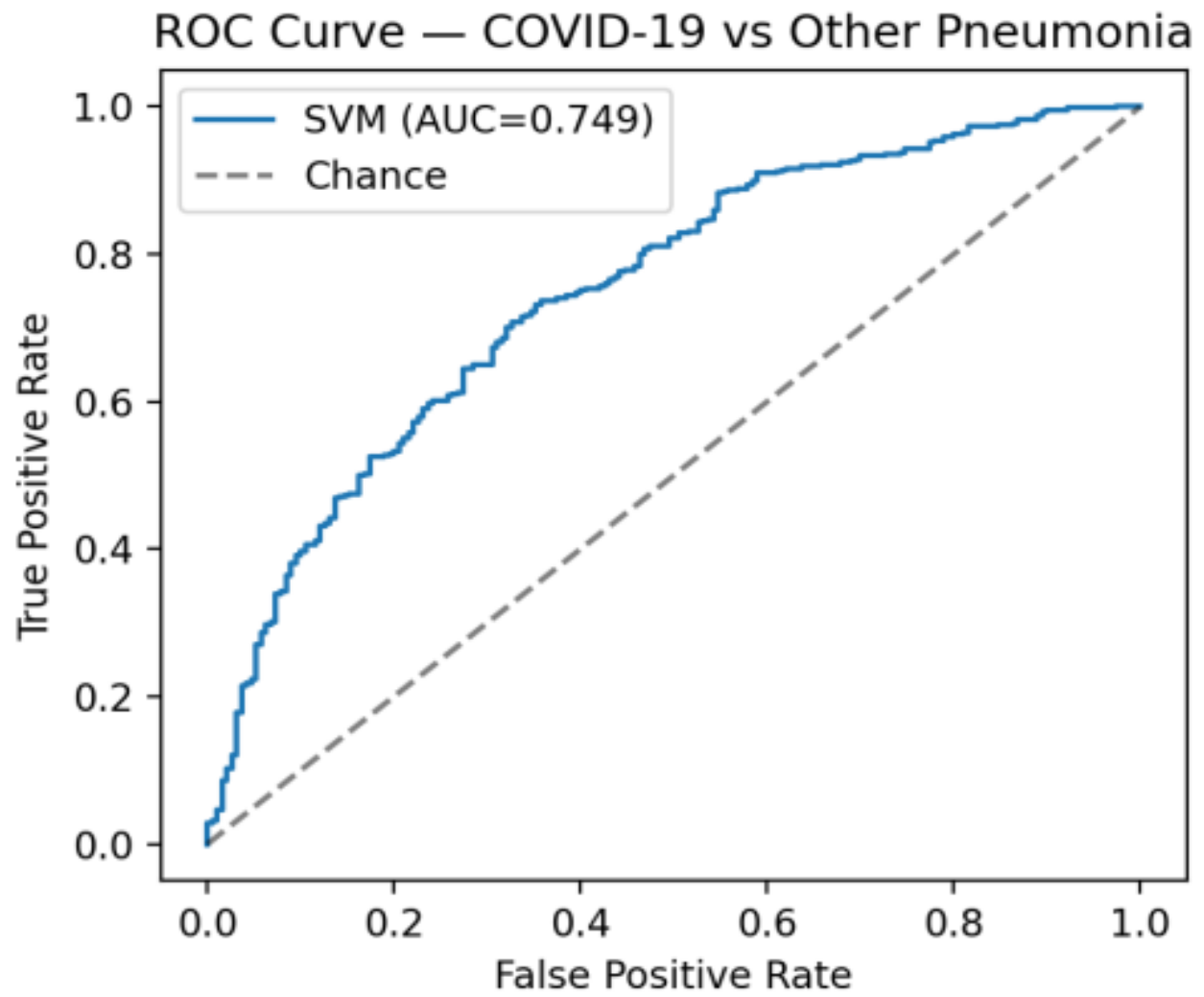

Fig. 2. Pooled ROC curve, SVM classifier, 5-fold CV (AUC = 0.755).

Recall was consistently higher than precision across models (e.g., random forest recall 0.985 vs. precision 0.743), indicating a bias toward predicting the majority COVID-19 class even with class-balanced weighting — a direct consequence of the limited number of other-pneumonia examples (190) and their heterogeneous etiologies (viral, bacterial, and fungal cases pooled into a single class).

## VI. Discussion and Limitations

### *A. Interpretation of Results*

The modest but real improvement over the majority-class baseline (75.4% vs. 71.6% accuracy; AUC 0.755 vs. 0.500) indicates that handcrafted texture and gradient features carry some discriminative signal for this task, consistent with the broader literature showing that COVID-19 pneumonia often presents with distinct ground-glass and peripheral opacity patterns. However, the effect size is small, and these results should not be interpreted as clinically meaningful diagnostic performance.

### *B. Limitations*

This study has several limitations that we report directly rather than obscure. First, the dataset is small (668 images, 408 patients) and drawn from heterogeneous published case reports rather than a systematic clinical screening population, which likely introduces selection bias. Second, the other-pneumonia class pools multiple distinct etiologies, which may reduce class coherence and classifier performance. Third, no independent external test set from a separate institution was available, so generalization cannot be assessed. Fourth, this pilot uses classical handcrafted features rather than learned representations; deep convolutional or transformer-based encoders may capture more discriminative patterns given sufficient data, but attempting to train such models on a dataset this size risks severe overfitting and was intentionally avoided here. Finally, no clinical validation with radiologists was conducted, and none of the results in this paper should be read as evidence of clinical readiness.

### *C. Toward a Multi-Modal Deep Learning Framework*

Based on this baseline and the reviewed literature, we outline — as a proposed direction for future work, not a completed contribution — a multi-modal architecture combining modality-specific convolutional-transformer encoders (e.g., an EfficientNet-style CNN stream paired with a Swin-style transformer stream, following the design principles in [5], [9], [10]) with a cross-attention fusion module integrating imaging and structured clinical metadata. Recent work on adapting large computer-vision foundation models to specialized domains through targeted fine-tuning [6] suggests a complementary path to training such encoders from scratch, particularly when labeled medical data remains scarce. Realizing this architecture responsibly would require: (1) access to a larger, multi-institutional dataset with proper IRB approval and documented data-sharing agreements; (2) an external, held-out test set from institutions not represented in training; (3) released code and model weights for reproducibility; and (4) a prospective reader study with radiologists before any claim of clinical utility. We view the pilot results in this paper as a necessary, honest first step establishing a reproducible baseline against which such a system could later be meaningfully compared.

## VII. Code and Data Availability

The COVID-19 Image Data Collection used in this study is publicly available at https://github.com/ieee8023/covid-chestxray-dataset under the terms specified by its maintainers [11]. Feature extraction, model training, and figure generation code for this study are available at https://github.com/2100030105/covid-pneumonia-xray-pilot-study. Providing this code is intended to allow independent verification of the reported results.

## VIII. Conclusion

This paper presented a small, fully reproducible pilot study for COVID-19 versus other-pneumonia classification using a publicly available chest X-ray dataset, classical texture and gradient features, and rigorous patient-level cross-validation. The best-performing model (SVM) achieved 75.4% accuracy and an AUC of 0.755, a modest improvement over a majority-class baseline. Rather than overstate these results, we use them as an honest empirical foundation to motivate a scoped, clearly-labeled proposal for future multi-modal deep learning work, contingent on access to larger, properly governed multi-institutional datasets and independent clinical validation.